# Quantum phase transition in magnetic nanographenes on a lead superconductor

Yu Liu[1†], Can Li[1†], Fu-Hua Xue[2†], Ying Wang[2], Haili Huang[3], Hao Yang[3], Jiayi Chen[1], Dan-Dan Guan[1,3], Yao-Yi Li[1,3], Hao Zheng[1,3], Canhua Liu[1,3], Mingpu Qin[1,3], Xiaoqun Wang[1,3], Deng-Yuan Li[2*], Pei-Nian Liu[2*], Shiyong Wang[1,3*], Jinfeng Jia[1,3*]

[1]*Key Laboratory of Artificial Structures and Quantum Control (Ministry of Education), Shenyang National Laboratory for Materials Science, School of Physics and Astronomy, Shanghai Jiao Tong University, Shanghai 200240, China*

[2]*Key Laboratory for Advanced Materials and Feringa Nobel Prize Scientist Joint Research Center, Frontiers Science Center for Materiobiology and Dynamic Chemistry, State Key Laboratory of Chemical Engineering, School of Chemistry and Molecular Engineering, East China University of Science & Technology, 130 Meilong Road, Shanghai, 200237, China*

[3]*Tsung-Dao Lee Institute, Shanghai Jiao Tong University, Shanghai, 200240, China*

[†]These authors contributed equally to this work.
*Corresponding Authors: dengyuanli@ecust.edu.cn, liupn@ecust.edu.cn, shiyong.wang@sjtu.edu.cn, jfjia@sjtu.edu.cn

**Quantum spins, referred to the spin operator preserved by full SU(2) symmetry in the absence of the magnetic anistropy[1–3], have been proposed to host exotic interactions with superconductivity[4]. However, spin orbit coupling[5,6] and crystal field splitting[7,8] normally cause a significant magnetic anisotropy for *d/f*-shell spins on surfaces[6,9], breaking SU(2) symmetry and fabricating the spins with Ising properties[10]. Recently, magnetic nanographenes have been proven to host intrinsic quantum magnetism due to their negligible spin orbital coupling and crystal field splitting[10–12]. Here, we fabricate three atomically precise nanographenes with the same magnetic ground state of spin S=1/2 on Pb(111) through engineering sublattice imbalance in graphene honeycomb lattice[13]. Scanning tunneling spectroscopy reveals the coexistence of magnetic bound states and Kondo screening in such hybridized system[14]. Through engineering the magnetic exchange strength between the unpaired spin in nanographenes and cooper pairs, quantum phase transition from the singlet to the doublet state has been observed, in consistent with quantum models of spins on superconductors[1,14,15]. Our work demonstrates delocalized graphene magnetism host highly tunable magnetic bound states with cooper pairs, which can be further developed to study the Majorana bound states[16–19] and other rich quantum physics of low-dimensional quantum spins on superconductors[10,12].**

Superconductor-magnet hybridized systems bear fruitful quantum phases, such as magnetic bound states[5,9,20–27], Majorana zero modes[16–19,28] and heavy-fermion superconductivity[29,30], with implications for advancing strongly correlated physics and quantum technological applications[21,25,31]. The theory of classic spins on superconductors was independently worked out by Yu[32], Shiba[33] and Rusinov[34] in 1960s, predicting the presence of a pair of bound states inside the superconductor gap. However, a fully description of quantum spins on superconductors are too complicated to be strictly solved. In 1977, a correlated Kondo state and magnetic bound states have been predicated to coexist in systems of quantum spins on superconductors by Matsuura using approximation methods at the strong coupling limit[1]. Using numerical renormalization group (NRG) methods, quantitatively understanding at different coupling strength between quantum spins and cooper pairs has been realized. With increased magnetic coupling strength between a spin S=1/2 and cooper pairs, the ground state of the system transforms from the doublet state to the singlet ground state by a quantum phase transition at a critical coupling point of $k_B T_k \approx 0.3\Delta$ [2,3].

Scanning tunneling microscopy has been widely used to study magnetic impurities on superconductors due to its high spatial and energy resolutions. In 1997, Eigler and coworkers provided the first experimental evidence of magnetic bound states on a Nb superconductor[20]. In 2008, superconducting tip had been employed to detect magnetic bound states with much higher energy resolution[35], allowing for distinguishing different magnetic bound states of single atoms[25], molecules[5,14,22,23,27,36,37], and spin chains[28] on different superconductors. In 2011, competition between the quasiparticle excitation and the Kondo screening has been observed in a multi-channel S=1 molecule on lead[14]. Up to date, the magnetism of most studied impurities originates from highly localized *d/f* orbitals of transitional metals, where crystal field and spin orbit coupling cause a noticeable magnetic anisotropy on surfaces[5,7,8,22], hindering the study of intrinsic quantum spins interacting with cooper pairs. Recently, a new type of delocalized π electron magnetism has been realized in nanographenes[38–46], small pieces of graphene, by using advances of surface chemistry[42,45], allowing for precisely engineering of their magnetic properties down to the single-chemical-bond level[13,38]. The magnetism of nanographene is distinct from transitional metals in three aspects: i) the spin density is delocalized inside the molecule[38,47]；ii) the spin orbital coupling of graphene material is negligible; and iii) the magnetic ground state can be easily engineered by introducing sublattice imbalance according to Lieb's theorem[13,47]. Additionally, such a large size nanographene hosts negligible magnetic anisotropy on surfaces due to its reduced crystal field splitting, and thus can be viewed as ideal quantum spins on surfaces[10,12].

In this letter, we report the study of three atomically precise magnetic nanographenes on superconducting Pb(111) film by using low-temperature scanning tunneling microscopy (STM). Through STM tip induced atom manipulation, three different nanographenes with the same magnetic ground state of S=1/2 have been fabricated on Pb(111), which have different adsorption configurations and thus different magnetic exchange strength with cooper pairs. Scanning tunneling spectroscopy (STS) reveals the presence of a pair of magnetic bound states inside the superconducting gap with drastic different binding energy and particle-hole asymmetry depending sensitively on the adsorption details. In addition, we quantified the Kondo temperatures of different nanographenes on lead by quenching the superconductivity of STM tip and lead by applying an out-of-plane magnetic field. Such measurements reveal the formation of the singlet ground state with the coupling strength ($k_B T_k > 1.6\,\Delta$) and the doublet ground state with the coupling strength ($k_B T_k < 1.6\,\Delta$), with a quantum phase transition at the critical point of $k_B T_k \sim 1.6\,\Delta$[14,15]. Our observations directly confirm a quantum phase transition induced by varying magnetic exchange strength between the unpaired spin in nanographene and cooper pairs, providing a highly tunable platform for further exploring Majorana zero mode and other quantum physics in graphene-superconductor hybridized system[16,17,19].

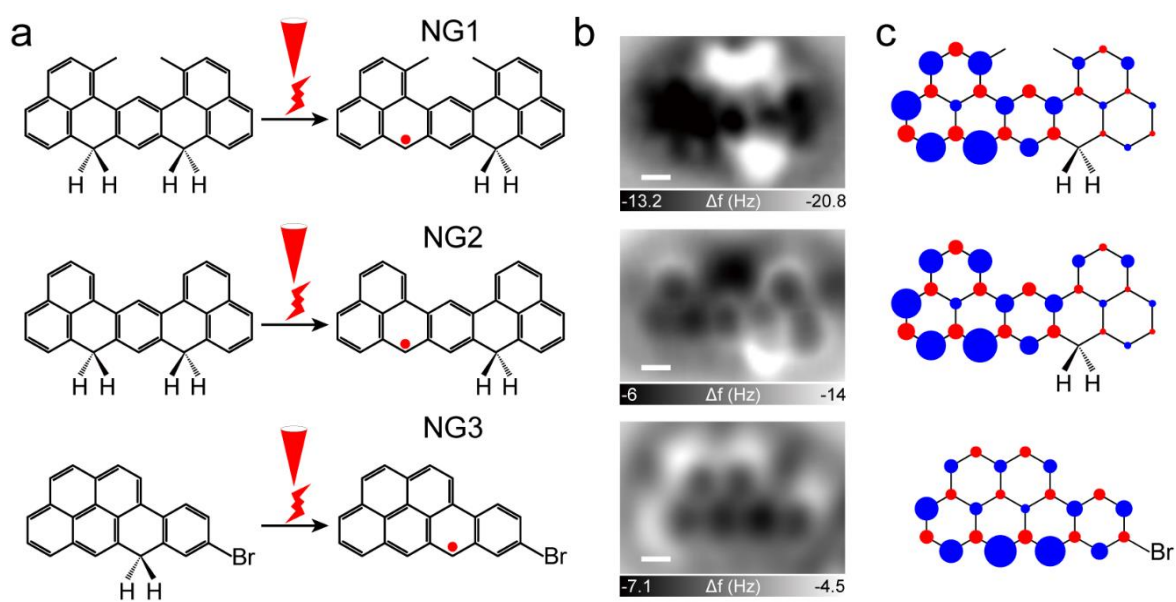

**Figure 1. Structural and magnetic properties of three S=1/2 magnetic nanographenes. a**, STM tip induced atom manipulation to generate magnetic nanographenes on surfaces. **b**, Nc-AFM frequency shift images ($f_0$ = 26 KHz, Oscillation amplitude of 80 pm) of nanographenes on Au(111). The methyl and sp3 carbon are resolved as bright protrusions due to an out-of-plane distortion. Scaler bar: 200pm.

**c**, Spin density distributions calculated by mean-field Hubbard model. Blue/red indicates spin up/spin down densities.

As shown in Figure 1, we synthesize three closed-shell nanographenes in solution and sublimate them onto a clean Au(111) surface under ultra-high vacuum conditions. Through STM tip induced atom manipulation[38], one hydrogen at the $sp^3$ carbon site can be controllably dissociated, which is accompanied by introducing one unpaired electron into the π electron topology of nanographene, giving rise to the S=1/2 magnetic ground state. Hereinafter, we denoted the three magnetic nanographenes as NG1, NG2 and NG3 for convenience. Their chemical structures have been resolved by non-contact atomic force microscopy functionalized with a carbon monoxide molecule. The presence of methyl groups and/or $sp^3$ carbons in NG1 and NG2 induces a local out-of-plane distortion on Au(111), giving rise to different scattering potential with conduction electrons of supporting substrates. Mean-field Hubbard (MFH) calculations suggest the presence of a net spin of S=1/2 in all the three nanographene with delocalized spin density distributions (Fig.1c and Supplementary figure 1), which is confirmed by STS measurements showing a pronounced zero-bias Kondo resonance on Au(111) (Supplementary figure 2)[48].

The magnetic bound states of these nanographenes on superconducting Pb(111) film have been detected by STS measurements. To explore the sub-millielectron energy scale in our system, we use a superconducting tip to suppress the thermal broadening energy near the Fermi level. High resolution dI/dV spectra resolve a pair of bound states inside the superconducting gap when positioning the tip over certain positions of NG1 as depicted in Fig. 2a. The binding energy for NG1 is well away from the superconducting gap edge with a value of $V_b=\pm1.3$meV. The peak intensity of in-gap bound states changes drastically when locating STM tip at different positions of NG1, but the binding energy keeps nearly constant. Constant-height STM images taken at the binding energy resolve the spatial distribution of in-gap states, which exhibits nodal patterns inside a NG1 and agrees with the MFH simulated density of states map using the singly occupied molecular orbital (Fig. 2c). Such delocalization character has also been confirmed by performing dense dI/dV spectra along the marked line in Fig. 2c (Fig. 2g). Similar observations have been made on NG2 and NG3 on Pb(111), but with slightly different binding energy of $V_b=\pm1.65$meV and $\pm1.54$meV, respectively. As highlighted by red dished line in Fig. 2a, the particle-hole symmetry of magnetic bound states is broken by showing different peak weights above and below the Fermi level. For the NG1, the hole-like peak is slightly stronger, while the particle-like peak is stronger for the NG2 and NG3. In the classic-spin model, the quantum mechanical properties of spins are completely neglected, and spin impurities are simply treated as a local magnetic field. Such classic model predicates the bound states appear in pairs with equal weights at positive and negative energies. The observation of particle-hole asymmetry indicates the internal degree of spins should be taken into account to fully capture the behind physics.

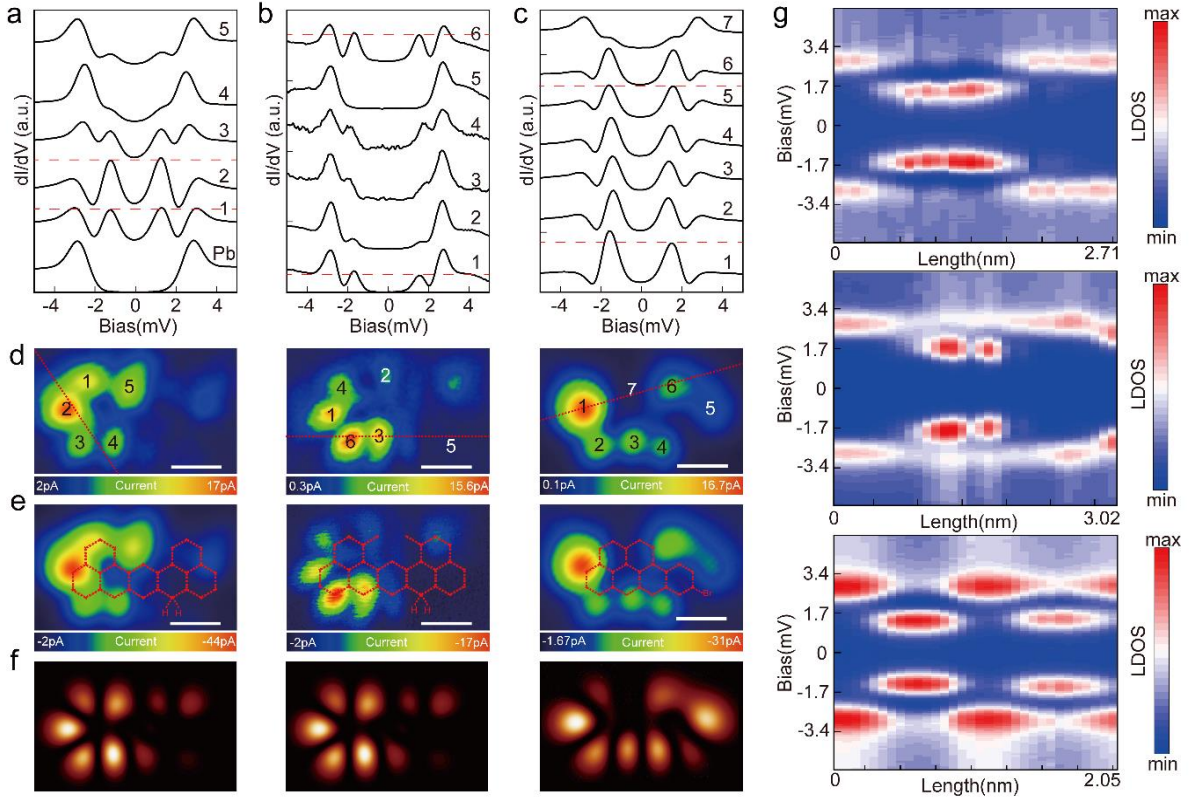

**Figure 2**. **Spatial distribution of the magnetic bound states in nanograhenes. a-c**, dI/dV spectra taken at the numbers marked in d using a superconducting tip. Setpoint: I=100pA, $V_{mod}$=20μV. A pair of in-gap magnetic bound states have been observed with intensities proportional to the spin density distributions of NGs. The inset red line indicates different spin polarization intensity at different sites of the orbital. **d-e**, Constant-height current images taken at the energy of the bound states. (Scaler bar: 500pm) The inset red line shows the line for dense spectrum. **f**, Simulated LDOS mappings using the singly occupied molecular orbital based on the mean-field theory. **g**, The dense dI/dV spectra through the nanographene. Setpoint: I=100pA, $V_{mod}$=20μV.

In the quantum-spin model, a well-explored case is a spin impurity which is antiferromagnetically exchanged with the reservoir bath[1]. The magnetic bound states correspond to a Bogoliubov quasiparticle excitation from ground state to the first excited state. The local Coulomb interaction around the impurity breaks particle-hole symmetry and leads to a spin-polarized magnetic bound states manifested as asymmetric spectral weights in tunneling spectra (Fig. 3a). For weak coupling, the ground state is an underscreened doublet state, and the impurity bound states correspond to the excitation from the doublet ground state to the singlet excited state with dominated hole-like spectral weight; For strong coupling, the ground state is a Kondo singlet state, and the impurity bound states correspond to the excitation from the singlet ground state to the doublet excited state with dominated particle-like spectral weight. This is an example of a quantum phase transition which occurs for a level crossing that depends on the energy scale of the kondo screening between impurity and cooper pairs (Fig. 3b). With increased strength of Kondo effect, the binding energy first decreases from Δ to zero, and then increased to Δ with a quantum phase transition near $k_B T_k \sim \Delta$. To systematically investigate the energy and particle-hole asymmetry of magnetic bound states, we measured 20 nanographenes with different adsorption configurations. The spectral features of magnetic bound states depend sensitively on adsorption details. Fig 3d shows several representative tunneling spectra obtained with a superconducting tip at 4.3 K and 1.1 K. Due to the convolution of tip and sample states, there are several in-gap peaks in dI/dV spectra originating from tip and sample density of states induced by thermal excitations. To remove the influence of superconducting density of states from the tip, we numerically deconvolute the raw data and get the authentic sample density of states, which reflects the intrinsic binding energy and the particle-hole asymmetry of the quasiparticle excitations (Supplementary figure 3). Fig. 3e shows the deconvoluted tunneling spectra with binding energy ranging from -0.82Δ to +0.4Δ, where the negative sign indicates the stronger particle-like peak and positive the stronger hole-like peak. A crossing of the magnetic bound states accompanied by the particle-hole asymmetry inversion is observed, confirming the quantum phase transition from the doublet to the singlet ground states as illustrated in Fig. 3a-b.

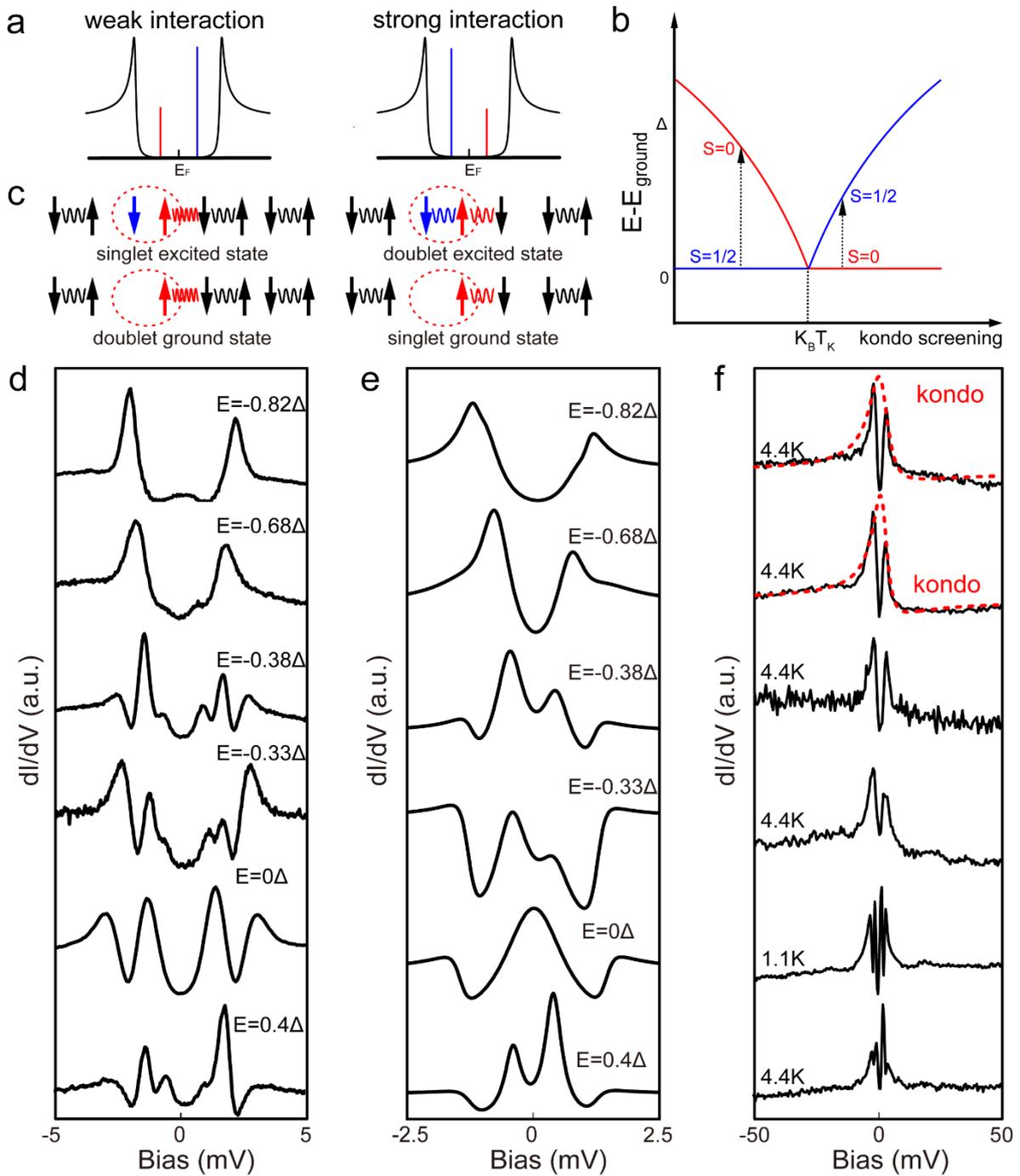

**Figure 3. Quasiparticle excitations with different binding energy and emergence of quantum phase transition. a**, Schematic diagram of the two coupling regimes of the quantum spin 1/2 on a superconductor. The weak interaction corresponds to the singlet ground state and the strong interaction corresponds to the doublet ground state. **b**, Quasiparticle excitation diagram for an unpaired spin on a superconductor. A quantum phase transition occurs for a level crossing that depends on the energy scale of the kondo screening between the spin impurity and cooper pairs. **c**, dI/dV spectra taken on five characteristic nanographenes using a superconducting tip. Setpoints: I=100pA, $V_{mod}$=10μV. **e**, Deconvoluted dI/dV spectra showing the intrinsic electronic properties barely from the sample. **f**, dI/dV spectra with a larger energy window for the same set of NGs as in **d**. Inset red dashed lines shows the envelope of the Kondo resonance. Setpoints: I=100pA, $V_{mod}$=50μV. All the spectra are shifted for clarity.

The magnetic interaction between the unpaired spin and itinerant electrons outside the superconducting gap has been probed by taking dI/dV spectra using a larger energy window. For strong coupling, a Frota-shape envelope has been observed in larger scale dI/dV spectra as marked by the dashed red lines, suggesting the unpaired spin has been screened by normal itinerant electrons outside the superconducting gap giving a Kondo singlet ground state (Fig. 3g). In the case of weak coupling, it is not possible to form a Kondo singlet since the absence of states in the vicinity of the Fermi level (Kondo peak width will be smaller than the superconducting gap), with a doublet ground state corresponding to an unscreened spin. The coexistence of the in-gap bound states and the Kondo resonance at strong coupling regime manifests the quantum spin behavior of the π magnetism in nanographenes on Pb(111), in consistent with numerical renormalization group calculations[1–3].

We quantified the magnetic exchange strength between nanographene and Pb(111) by measuring dI/dV spectra under an out-of-plane magnetic field, which can quench both the superconductivity of the tip and the lead substrate. Figure 4a shows dI/dV spectra under different magnetic field taken on a NG1 molecule. With increased magnetic field, the bound states gradually disappear, and a Kondo resonance emerges at Fermi level once the superconductivity is quenched. The peak width of kondo resonance reflects the energy scale ($K_B T_K$) of magnetic exchange coupling strength. The Kondo temperature $T_K$ can be estimated by using the relation $K_B T_K \approx 0.686 \Gamma$ ($\Gamma$ is the fitted Frota peak width) if the experimental temperature is well below the Kondo temperature[26]. As shown in Fig. 4b, Kondo temperatures ranging from 13K to 95 K have been obtained, well above the experimental temperature. Figure 4c depicts the relation between the binding energy E/Δ and the coupling strength $k_B T_k /\Delta$. The properties of magnetic bound states, including energy position and spectral weight, depends crucially on coupling strength with a level crossing near $K_B T_K \sim 1.6\Delta$, confirming a quantum phase transition between the doublet and the singlet ground state.

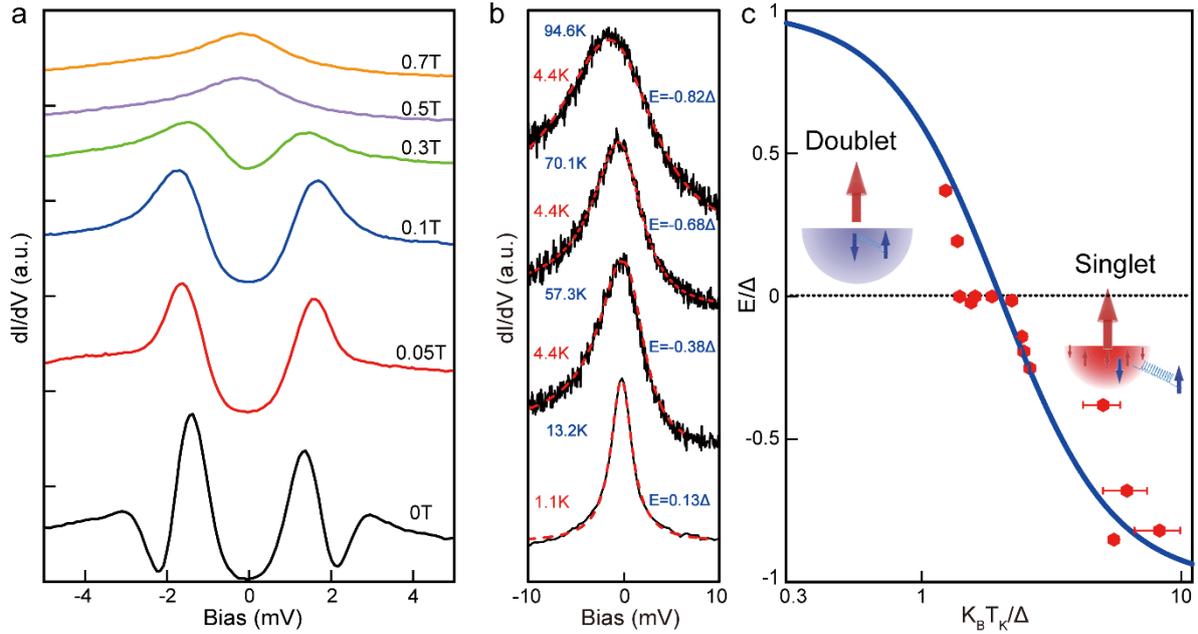

**Figure 4. Relation between the energy scale of the kondo screening and the binding energy of magnetic bound states. a**. Out-of-plane magnetic-field dependent dI/dV spectra taken at the same position of a NG1. The lines are shifted vertically for clarity. **b**. The Kondo resonance spectra of different nanographenes with different binding energy of magnetic states (black lines). The dashed red lines are fitted curves using a Frota function. **c**. Relation between the energy scale of Kondo screening and the binding energy of magnetic bound states. The inset diagram illustrates the many-body ground state of the hybridized system.

In summary, we directly observe the quantum phase transition from the doublet to the singlet ground state by varying the magnetic exchange strength between an unpaired spin in nanographene and cooper pairs. The nanographene-superconductor hybridized systems represent a rich platform for further exploring quantum physics between quantum spins and cooper pairs due to their highly tunability and π-electron delocalization. For example, the many-body quasiparticle excitation of intrinsic Heisenberg spin chains on superconductors[10,12,49] can be realized by choosing appropriate precursors[38,42]; The multi-channel quasiparticle excitation of high-spin nanographenes on superconductors can be explored[2,14,49]. This kind of highly tunable magnetism may be used for fabricating a Qubit based on the spin-singlet bound states, or even the topological Qubit based on Majorana zero-modes in pure carbon-based systems[31,50].

**Methods**

Sample preparation and characterization were carried out using a commercial low-temperature Unisoku Joule-Thomson and a USM-1300 scanning probe microscopy under ultra-high vacuum conditions (3x10$^{-10}$ mbar). The lead films were grown on clean highly oriented pyrolytic graphite (HOPG) with tens of nanometers thick. The nanographenes were deposited on Pb(111) and Au(111) holding at room (300K) or helium temperature (<10K). Afterwards, the sample was transferred to a cryogenic scanner at 1 K or 4.9 K for characterization.

STM images were obtained by maintaining a constant-current sweep mode for the topography and a constant-height mode for the orbital mapping. With feedback loop opened, the differential tunneling conductance dI/dV spectra were measured using a standard lock-in technique with a small modulation voltage $V_{mod}$ = 50 ueV at the frequency $f$=589Hz. To increase AFM resolution, CO molecule is picked up from Au(111) to the apex of a tungsten tip. A quartz tuning fork with resonant frequency of 26 KHz has been used to realize nc-AFM measurements. The superconducting tip was fabricated by indenting the tip into the lead films to a depth of several nanometers so that a superconducting cluster was picked up.

The tight binding (TB) calculation of the STM images were carried out in the C $2p_z$-orbital description by numerically solving the Mean-Field-Hubbard Hamiltonian with nearest-neighbor hopping:

$$\hat{H}_{MFH} = \sum_{<ij>,\sigma} -t_{ij} c^+_{i,\sigma} c^-_{j,\sigma} + U \sum_{i,\sigma} \langle n_{i,\sigma} \rangle n_{i,\bar{\sigma}} - U \sum_i \langle n_{i,\uparrow} \rangle \langle n_{i,\downarrow} \rangle$$

with $t_{ij}$ is the nearest-neighbor hopping term depending on the bond length between C atoms, and $c^+_{i,\sigma}$ and $c^-_{j,\sigma}$ denoting the spin selective ($\sigma = \{\uparrow, \downarrow\}$) creation and annihilation operators on the atomic site i and j, U the on-site Hubbard parameter (with U=3.5eV used here), $n_{i,\sigma}$ the number operator and $\langle n_{i,\sigma} \rangle$ the mean occupation number at site i. Numerically solving the model Hamiltonian yields the energy Eigenvalues $E_i$ and the corresponding Eigenstates $\alpha_{i,j}$ (amplitude of state i on site j) from which the wave functions are computed assuming Slater type atomic orbitals:

$$\psi_i(\vec{r}) = \sum_j \alpha_{i,j} \cdot (z - z_j) \exp(-\zeta |\vec{r} - \vec{r}_j|)$$

with $\zeta$=1.625 a.u. for the carbon $2p_z$ orbital. The charge density map $\rho(x,y)$ for a given energy range [$\varepsilon_{min}, \varepsilon_{max}$] and height $z_0$ is then obtained by summing up the squared wave functions in this chosen energy range.

$$\rho(x,y) = \sum_{i, \varepsilon_i \in [\varepsilon_{min}, \varepsilon_{max}]} \psi_i^2(x, y, z_0)$$

Constant charge density maps are taken as a first approximation to compare with experimental STM images.

**Acknowledgements**

We thank Prof. Roman Fasel, Dr. Yalei Lu and Dr. Chenxiao Zhao for fruitful discussions. S. W. acknowledges the financial support from National Key R&D Program of China (No. 2020YFA0309000), the National Natural Science Foundation of China (No. 11874258, No. 12074247), the Shanghai

Municipal Science and Technology Qi Ming Xing Project (No. 20QA1405100), Fok Ying Tung Foundation for young researchers and SJTU (No. 21X010200846). This work is also supported by the Ministry of Science and Technology of China (Grants No. 2019YFA0308600, 2016YFA0301003, No. 2016YFA0300403), NSFC (Grants No. 21925201, No. 22161160319, No. 20ZR1414200, No. 21925201, No. 11521404, No. 11634009, No. 92065201, No. 11874256, No. 11790313, and No. 11861161003), the Strategic Priority Research Program of Chinese Academy of Sciences (Grant No. XDB28000000) and the Science and Technology Commission of Shanghai Municipality (Grants No. 22QA1403000, No. 20ZR1414200, No. 2019SHZDZX01, No. 19JC1412701, No. 20QA1405100) for partial support.

**Author Contributions**

S.W. and J.J. conceived the experiments. Y.L., and C.L. performed the STM experiments. F.X., and Y. W. synthesized the molecular precursors under the supervision of D.L. and P. L. Data interpretations were conducted by J.C., H.Y., H.H., D.G., Y.L., H.Z., C.L., M.Q., X.W., Y.L., C.L. J.J. and S.W. The paper was written by S.W., Y.L., and C.L. All authors commented on the manuscript at all stages.

**Competing financial interests**

The authors declare no competing financial interests.

**Data availability.**

The datasets generated and/or analysed during the current study are available from the corresponding author on reasonable request.

**Code availability.**

The tight-binding calculations were performed using a custom-made code on the Matlab platform. Details of this tight-binding code can be obtained from the corresponding author on reasonable request.